\documentclass[PRL,aps,twocolumn,showpacs,preprintnumbers,amsmath,amssymb,shperscriptaddress,nofootinbib]{revtex4-1}
\usepackage[dvips]{graphicx}
\usepackage{epsf}
\usepackage{amsmath}
\usepackage{amssymb}
\usepackage{amsfonts}
\usepackage{times}
\usepackage[usenames]{color}
\usepackage[normalem]{ulem}
\usepackage[T1]{fontenc}
\usepackage[dvipsnames]{xcolor}
\usepackage{floatrow}

\usepackage{graphicx}
\usepackage{dcolumn}
\usepackage{bm}
\usepackage{color}

\begin{document}

\title{Unveiling the wave nature of gravitational-waves with simulations}

\author{Jian-hua He}
\email[Email address: ]{hejianhua@nju.edu.cn}
\affiliation{School of Astronomy and Space Science, Nanjing University, Nanjing 210093, P. R. China}
\affiliation{Key Laboratory of Modern Astronomy and Astrophysics (Nanjing University), Ministry of
Education, Nanjing 210093, China}

\begin{abstract}
We present the first numerical simulations of gravitational waves (GWs) passing through a potential well generated by a compact object in 3-D space, with a realistic source waveform derived from numerical relativity for the merger of two black holes. Unlike the previous work, our analyses focus on the time-domain, in which the propagation of GWs is a well-posed "initial-value" problem for the hyperbolic equations with rigorous rooting in mathematics and physics. Based on these simulations, we investigate for the first time in realistic 3-D space how the wave nature of GWs affects the speed and waveform of GWs in a potential well. We find that GWs travel faster than the prediction of the Shapiro time-delay in the geometric limit due to the effects of diffraction and wavefront geometry. As the wave speed of GWs is closely related to the locality and wavefront geometry of GWs, which are inherently difficult to be addressed in the frequency-domain, our analyses in the time-domain, therefore, provide the first robust analyses to date on this issue based on solid physics. Moreover, we also investigate, for the first time, the interference between the incident and the scattered waves (the "echoes" of the incident waves). We find that such interference makes the total lensed waveforms dramatically different from those of the original incident ones not only in the amplitude but also in the phase and pattern, especially for signals near the merger of the two back holes. 
\end{abstract}

\maketitle
{\bf Introduction.}
If GWs are far away from the source, GW signals $h_{\mu\nu}(\vec{x},t)$ can be treated by linear theory. The equations governing GWs are analogous to those of Maxwell's equations in the vacuum. GWs $h_{\mu\nu}(\vec{x},t)$, therefore, exhibit wave effects, resembling those of the electromagnetic fields. One prominent property of the traveling waves is that  they are {\it retarded} signals (see e.g. Ref.~\cite{GR_steven_weinberg})
\begin{equation}
h_{\mu\nu}(\vec{x},t)=\frac{4G}{c^4}\int \mathrm{d}^3x'\frac{S_{\mu\nu}(\vec{x}',t-|\vec{x}-\vec{x}'|/c)}{|\vec{x}-\vec{x}'|}\,,\label{hwave}	
\end{equation}
where $S_{\mu\nu}(\vec{x}',t)$ is the energy-momentum tensor of the source. The {\it retarded} signals suggest that GWs travel with a finite speed. This finiteness leads to a fundamental property of GWs that the effects of GWs are local: they only affect regions that have causal connections. 

In addition to the locality, another less noticed but important property of GWs is that in 3-D space (or odd dimensions) they obey the strong Huygen's principle. Unlike in the 2-D case (or even dimensions), 3-D waves have clear wavefronts and the signals are "Sharp". A perturbation at a point $\vec{x}$ is visible at another point $\vec{x}'$ exactly at the time $t = |\vec{x} - \vec{x}'|/c$ but not later (see Eq.~(\ref{hwave})). If GWs are time-finite, the sharpness of wavefronts means that there will be clear leading and trailing wavefronts for wave-zones (2-D waves do not have a "Sharp" trailing wavefront). Huygen's principle also states that every point on the wavefront can be considered as the source of secondary wavelets and the next wavefront is determined by the envelope of those secondary wavelets. Since the wave speeds of those secondary wavelets are finite, the propagation of GWs is local, which is not affected by remote boundaries that do not have causal connections.  
 
Despite the importance of locality and Huygen's principle of GWs, they have not been properly addressed in the literature, as most of the previous works focus on the frequency-domain~\cite{PhysRevD.34.1708,Meena:2019ate,Deguchi,Schneider,Ruffa_1999,DePaolis:2002tw,Takahashi:2003ix,1999PThPS.133..137N,Suyama:2005mx,Christian:2018vsi,DOrazio:2019ens,Zakharov_2002,Dai:2017huk,Liao:2019aqq,Jow:2020rcy,Macquart:2004sh,PhysRevLett.80.1138,Dai:2018enj,PhysRevD.90.062003,Yoo:2013cia,Nambu:2019sqn}. Contrary to the common intuition that the time-domain and frequency-domain are equivalent, the frequency-domain, indeed, has fundamental limitations to address some time-domain issues, especially for the locality of GWs. This is because the Fourier transform assumes infinite spacetime, which requires global information, while the propagation of GWs is local, which is only valid in part of the spacetime. Moreover, after the Fourier transforms, the wave equations in the frequency-domain become elliptic (e.g. Helmholtz equations), which are specific to the "boundary-value" problems~\cite{nla:cat-vn1414651}. The wave functions, in this case, are no longer determined by initial conditions but by boundary conditions. So to get equivalent results, the boundary conditions in the frequency-domain have to be consistent with those in the time-domain, which, indeed, only works for a few limited cases. If the wave functions in the time-domain themselves are for "boundary-value" problems, such as steady-state (e.g. Ref.~\cite{Peters}) or stationary waves (e.g. Ref.~\cite{Suyama:2005mx}), the boundaries in the frequency-domain can be consistently set. However, if the wave functions in the time-domain are for "Cauchy" problems, such as the propagation of GWs, we have to know the evolution of the wave functions in the first place. For simple cases, such as finite wavelets traveling in free space, the future behaviors of the wavelets at infinity can be obtained using the Green's functions. The boundaries in the frequency-domain, in this case, can be consistently set as the radiation boundaries, like those in the Kirchoff's diffraction theory for waves in free space~\cite{Kirchoff}. However, if for a more complicated problem, such as waves traveling in a potential well, the wave-zones may have complex geometry in 3-D space (may not even continuous) due to the locality of waves, it is difficult to obtain the evolution of the wave functions in the first place and, hence, difficult to set consistent boundaries in the frequency-domain. In these cases, GWs should be studied in the time-domain directly.  
 
Besides, the propagation of GWs in the time-domain, itself, is a well-posed "initial-value" ("Cauchy") problem for hyperbolic equations with rigorous rooting in mathematics (see e.g. Ref.~\cite{grossmann2007numerical}). The wave functions in the time-domain are also explicitly related to local physical laws, such as the conservation of energy-momentum. Although it is usually difficult to get analytical solutions in the time-domain due to the complexity of the wave equations, because of the locality, the propagation of GWs can be effectively studied using numerical simulations even with a limited simulation domain.  

Here, we report the first time-domain simulations of GWs passing through a compact object using the modern state-of-the-art numerical technique. Based on our simulations, we present a robust analysis of how the wave effects, such as wavefront, interference, diffraction, and locality, affect the speed as well as the waveforms of GWs when they travel in a potential well. 

\begin{figure}
\includegraphics[width=\linewidth]{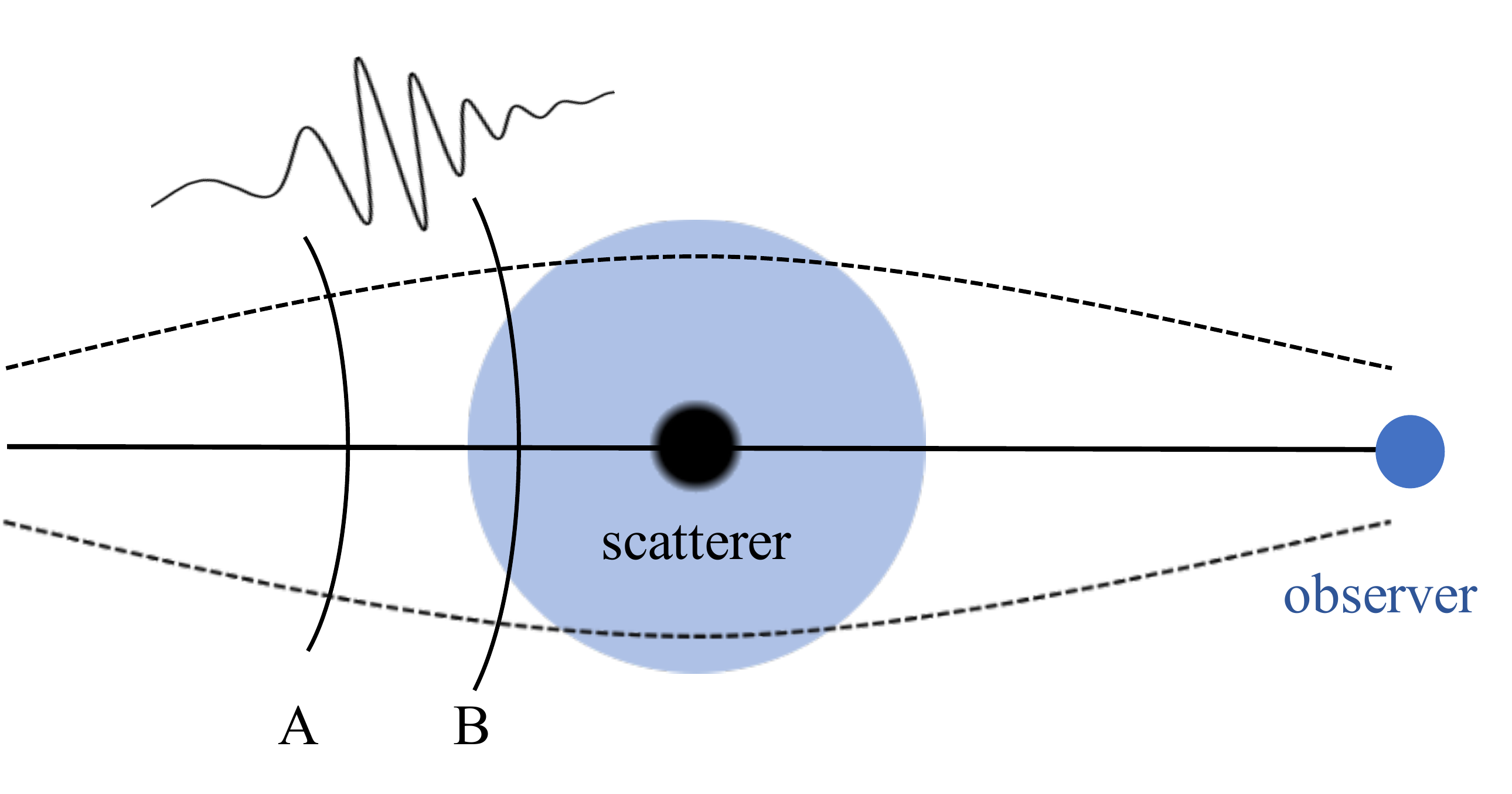}
\caption{The schematic of GWs scattered/lensed by a compact object. The scattered GWs signals are significant only in a short period near the merger of black hole binaries. If viewed in the Universe, these signals are geometrically confined in a thin-shell between the wavefronts $\rm A$ and $\rm B$. The wave-zone of the signals in 3-$\rm D$ space, therefore, has a complex geometry relative to the scatterer/lens.  \label{Figone}}
\end{figure}

{\bf Simulation setups and the input waveform.} The details of the numerical method used in this work are presented in our companion paper~\cite{He:2019orl}. Readers are referred to this paper for details. A notable feature of our numerical method is that we numerically solve the wave equations for the scattered waves $\delta h := h-\tilde{h}$, instead of the total wave function $h$
\begin{eqnarray}
c^2\nabla^2 \delta h -\frac{\partial^2}{\partial t^2}\delta h =-4c^2\psi \frac{\partial^2}{\partial t^2}\tilde{h} \,, \label{deltau}
\end{eqnarray}
where $c^2=1/(1-4\psi)$ is the effective wave speed in the presence of potential well $\psi$ and
$\tilde{h}$ is the original incident waves travelling in free space from the source, which satisfies 
\begin{align}
\nabla^2 \tilde{h} -\frac{\partial^2}{\partial t^2}\tilde{h} &=0\quad.
\end{align}
The above equation has a very simple analytical solution
\begin{equation}
\tilde{h}(x,t)=\left \{
	\begin{aligned} 
	&\frac{Q(t-\frac{r}{c})}{4\pi r} \quad &t\ge r/c\\
	&0\quad &t<r/c
	\end{aligned}
	\right. \quad, \label{sourcewf}
\end{equation}
where $r$ is the radial distance of the wavefront relative to the source as illustrated in Fig.~\ref{Figone}. $Q(t-r/c)$ is the waveform function. The term $t-r/c$ indicates that the waves are a {\it retarded} signals with a wave speed of $c$.

\begin{figure}
\includegraphics[width=\linewidth]{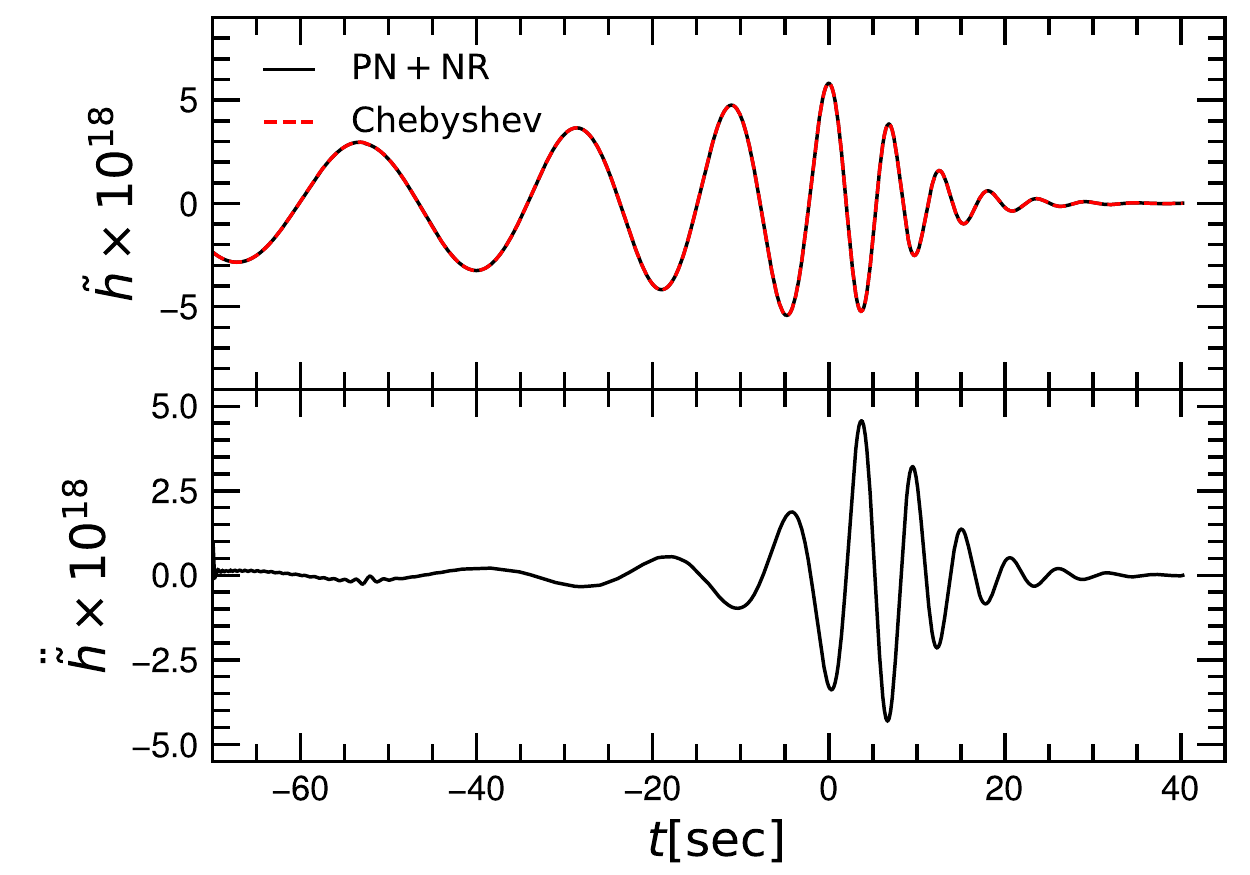}
\caption{The strains of GWs at the scatterer/lense as a function of time (black solid lines in the upper panel). The source is assumed to be generated by GW radiation from equal-mass non-spin black hole binaries located $100\,{\rm Mpc}$ away from the scatterer/lense. The waveforms are derived from numerical relativity and then matched smoothly to Post-Newtonian inspiral data with $3.5$ PN-accurate in phase and $2.5$ PN-accurate in amplitude. The zero-point of the time axis is chosen at the epoch when the two black holes start to merge. As the source term in Eq.~(\ref{deltau}) is related to the second derivative of the incident waveform, the waveform is first fitted by the Chebyshev polynomials up to $200$ orders (red dashed lines in the upper panel) and then the second derivative waveforms are obtained from the derivative of the Chebyshev polynomials (black solid lines in the lower panel).\label{strains}}
\end{figure}

In this work, the waveform $Q(t)$ is taken as GW radiation from equal-mass non-spin black hole binaries with a total redshifted mass of $M=10^5M_{\odot}$. The waveform is derived from numerical relativity~\cite{Baker:2007fb} and then matched smoothly to Post-Newtonian inspiral data with $3.5\, {\rm PN}$-accurate in phase and $2.5\, {\rm PN}$-accurate in amplitude (see Ref.~\cite{Blanchet:2013haa} for a review). The source is located $100\,{\rm Mpc}$ away from the scatterer/lens. For simplicity, we assume that the binary plane is edge-on. The inclination is $90$ degrees. As such, only $h_{+}$ mode is non-vanishing. Figure~\ref{strains} shows the strains at the position of scatterer/lense as a function of time (black solid lines in the upper panel). The zero-point of the time axis is chosen at the epoch when the two black holes start to merge. 

Moreover, since the source term $-4c^2\psi \frac{\partial^2}{\partial t^2}\tilde{h}$ in Eq.~(\ref{deltau}) is related to the second derivative of the waveforms $\frac{\partial^2}{\partial t^2}\tilde{h}$ rather than the waveforms directly, to accurately get the second derivative, we first fit the waveform by Chebyshev polynomials up to $200$ orders (red dashed lines in the upper panel of Fig.~\ref{strains}) and then obtain the second derivative from the derivative of the Chebyshev polynomials (black solid lines in the lower panel of Fig.~\ref{strains}). Compared with the original waveforms, the second derivative waveforms are significant only near the merger of black hole binaries. Even for massive black hole binaries as investigated in this work, the signals only last for $\sim 40\, {\rm secs}$. If viewed in the Universe, these signals are geometrically confined in a thin-shell. The wave-zone of the signals in 3-$\rm D$ space, therefore, has a complex geometry relative to the scatterer/lens.

As for the scatterer/lense, we choose the potential $\psi$ of scatterer/lense as generated by a homogeneous sphere
with a total mass of $M=10^5M_{\odot}$ and a Schwarzschild radius as $R_s = 2M$. The scatterer/lens is put at the center of the simulation domain. The box size of the simulation domain in this work is taken as $50\,{\rm sec}$ along one side. The observer is put at a distance of $r=43\,{\rm sec}$ to the left side surface of the simulation box long the direction of the incident waves as illustrated in Fig.~\ref{Figone}. In our simulation, we also choose the leading wavefront at $65\,{\rm sec}$ before the merger of the black holes. 

{\bf Energy conservation.}
As already noted in the introduction, unlike in the frequency-domain, the propagation of GWs in the time-domain is a "Cauchy" problem, which is inherently local and determined by local physical laws. So it is important to check whether waves in our simulations obey the laws of physics. One important test is the conservation of energy. A prominent feature of our simulations is that the scheme of integration for the time evolution is symplectic~\cite{He:2019orl}, which is called the {\it Crank-Nicolson Scheme} (also known as the implicit midpoint rule) with second-order accuracy. Although the scheme is implicit and time-consuming, it is inherently energy-preserving.

\begin{figure}
{\includegraphics[width=\linewidth]{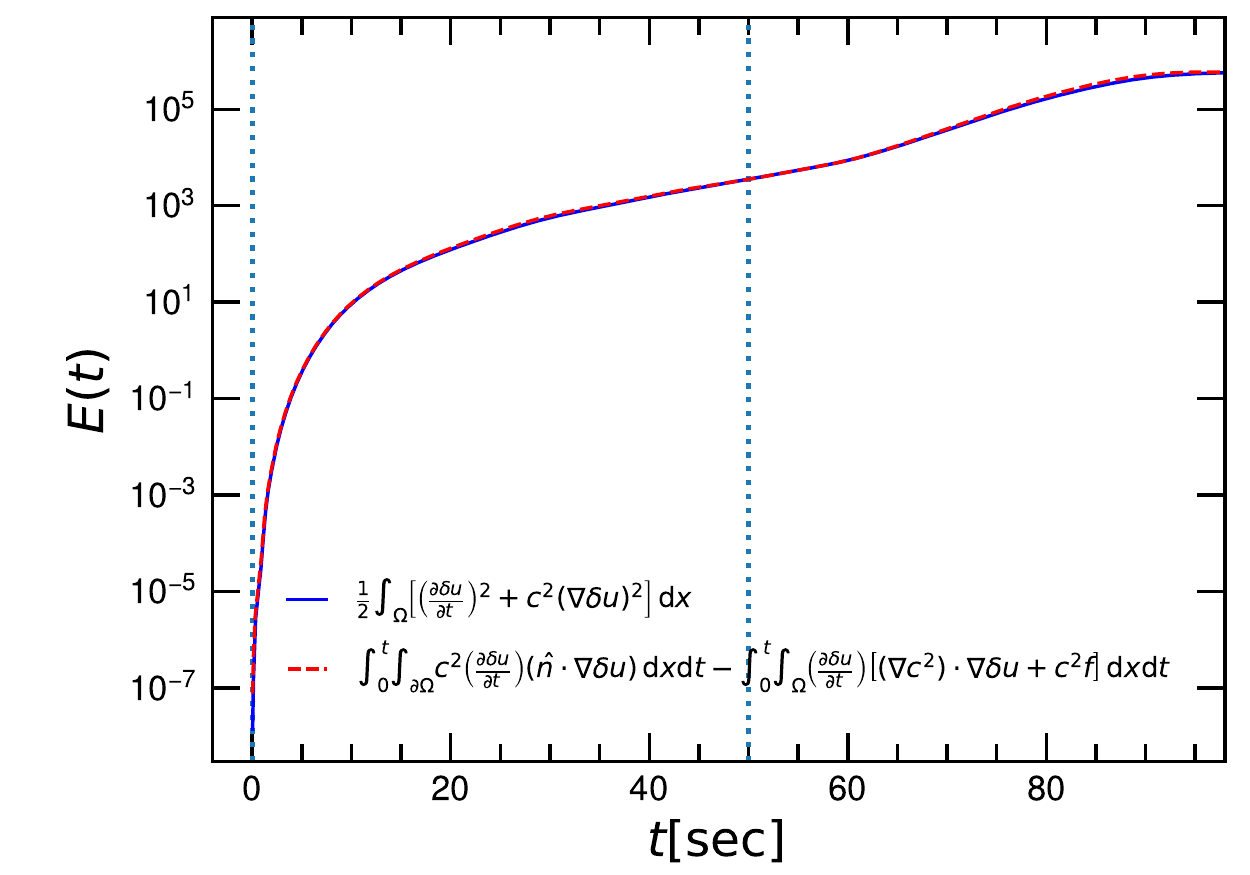}}
\caption{Energy conservation for the scattered wave $\delta h$. The blue lines represent the total energy of the scatted waves $\delta h$ measured in the simulation box. The red lines are the total energy generated by the source plus the energy passing through the surfaces of the simulation box. The dotted vertical lines indicate the epochs that the leading wavefront of the incident waves enters (left one) and leaves (right) the simulation box. \label{energy_conservation}}
\end{figure} 

Figure~\ref{energy_conservation} shows the total energy of the scatted waves $\delta h$ measured in the simulation box (blue solid lines) versus the energy generated from the source plus the energy passing through the surfaces of the simulation box. These two curves agree very well, meaning that the waves follow the law of the conservation of energy. The left vertical dashed line indicates the epoch when the leading wavefront of the incident wave enters the simulation box, for which we set as zero, while the right one indicates the epoch when the leading wavefront leaves the simulation box.

{\bf Diffraction and the wave speed of the scattered waves.}
From Eq.~(\ref{deltau}), it can be known that the scattered waves travel slower than those in the free space due to the presence of the potential well (see the wave speed $c^2=1/(1-4\psi)$). This phenomenon is known as the Shapiro time-delay~\cite{Shapiro}. From Eq.~(\ref{deltau}), it can also be known that the Shapiro time-delay does not explicitly depend on the frequency of the incident waves, as the wave speed of the scattered waves is solely determined by the potential $c^2=1/(1-4\psi)$. If the scattered waves are planar and $\psi$ is a constant, the results on the speed of waves are consistent with those from the geometric limit. However, if $\psi$ is not a constant and the wavefront has a complex geometry relative to the scatterer/lens, a great complexity comes in because waves far away from the center of the scatterer travel faster than those in close proximity to the center. In the geometric limit, these rays do not affect each other. However, due to Huygen's principle, every point on the wavefront can be considered as the source of secondary wavelets. These secondary wavelets can spread out in the forward direction. Thus, waves along faster rays may spread out to the slower ones, which, in turn, can boost the wave speed along with the slower rays. Moreover, because of the Shapiro time-delay and the geometry of the wavefront relative to the potential well, the secondary wavelets have different wave speeds along different directions. As a result, unlike in the free space (no gravitational potential), a "sharp" wavefront may no longer keep "sharp" but, instead, spread into complicated wave zones. 

\begin{figure}
{\includegraphics[width=\linewidth]{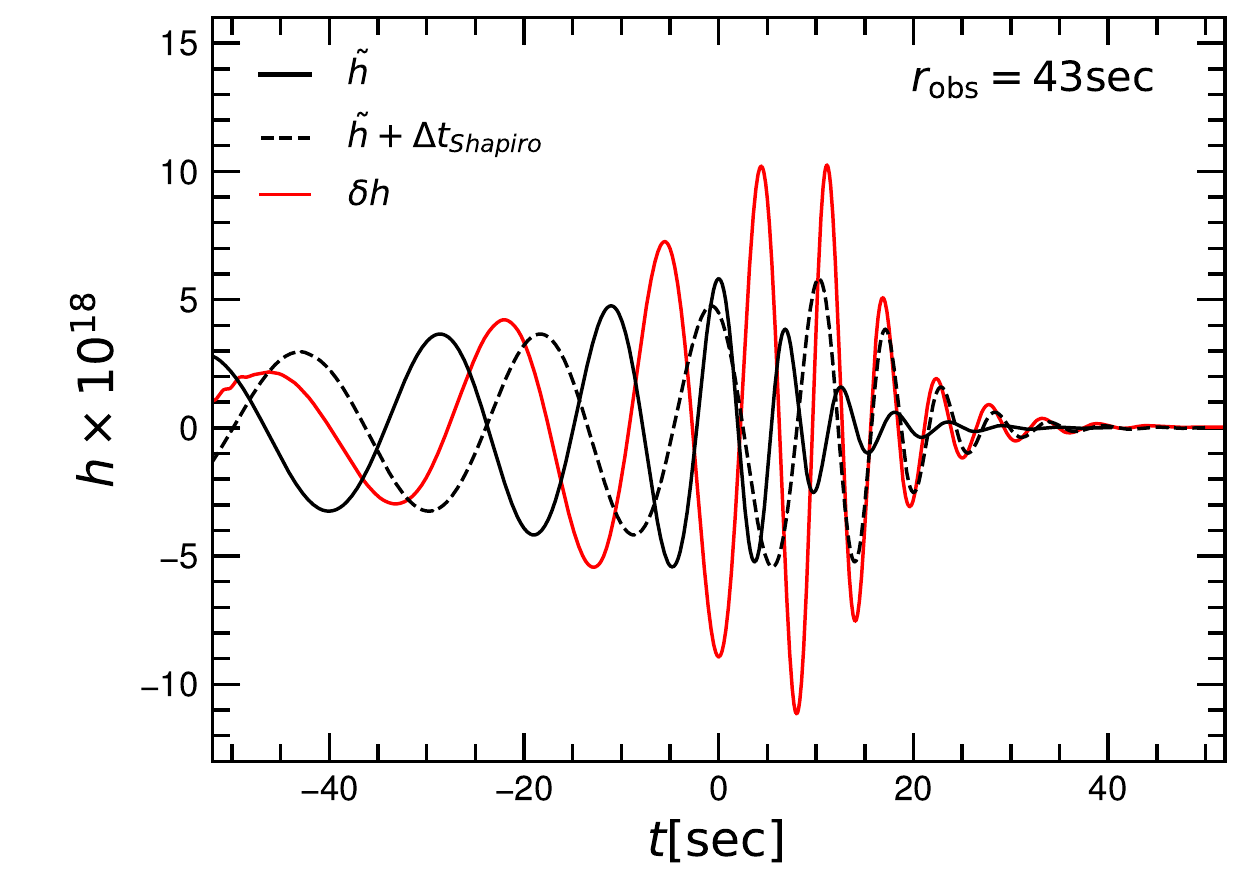}}
\caption{The effect of diffraction on the speed of GWs. Red and black solid lines are for the scattered waveforms $\delta h$ (red) and the original incident waveforms $\tilde{h}$ (black) at the position of the observer. The black dashed line shows the incident waveforms shifted by an amount of Shapiro time-delay $\Delta t_{\rm Shapiro}$ (see the text for details). The Shapiro time-delay are for the geometric limit, which gives the maximum possible time-delay. The difference is larger for the low-frequency signals (red and black dashes lines), which means that the low-frequency signals travel faster than those of the geometric limit due to the stronger diffraction effect. However, the difference is smaller for the high-frequency signals near the merger of black holes. The red curve in this case almost coincides with the dashed one, meaning that the time-delay in the high-frequency regimes is near the geometric limit due to the smaller diffraction effect. \label{wavespeed}}
\end{figure} 

Figure~\ref{wavespeed} compares the scattered waveforms $\delta h$ (red solid line) and the original incident waveforms $\tilde{h}$ (black solid line) at the position of the observer. As the observer is collimated, in this case, the only possible physical mechanism to delay the scattered waves is the Shapiro time-delay (no geometric delay). Next, we shall demonstrate that the change in the wave speed is due to diffraction. To do this, we show the incident waveforms shifted by the Shapiro time-delay $\Delta t_{\rm Shapiro}=2\int_{r_i}^{r_o}\psi dr$ as the black dashed line in Fig.~\ref{wavespeed}, where $r_i$ is the position of the surface of the simulation box and $r_o$ is the position of the observer. Note that the Shapiro time-delay is fortime-delay the geometric limits. The black dashed line in Fig.~\ref{wavespeed}, therefore, gives the maximum possible of the time-delay. Clearly, due to the effect of diffraction, which boosts wave speed, the time-delay in the scattered waveforms $\delta h$ is always less than the amounts due to the Shapiro time-delay (comparing the red curve with the dashed one). The difference is larger for the low-frequency signals, which means that the low-frequency signals travel faster than those of the geometric limit due to the stronger diffraction effect. However, the difference is smaller for the high-frequency signals near the merger of the black holes. The red curve in this case almost coincides with the dashed one, which means that the delay in the high-frequency signals is near the geometric limit due to the smaller diffraction effect.

{\bf Interference between the incident and the scattered waves.} 
As already noted, a "sharp" wavefront may no longer be "sharp" in a potential well. An equivalent way to understand this is that the incident wave can be considered as a source (see. Eq.~(\ref{deltau})), which triggers a series of scattered waves when passing through a potential well. Just like sound, these scattered waves can be considered as the "echoes" of the incident wave. Due to the interference, these "echoes" can blur the original signal, making it no longer "sharp". 

\begin{figure}
{\includegraphics[width=\linewidth]{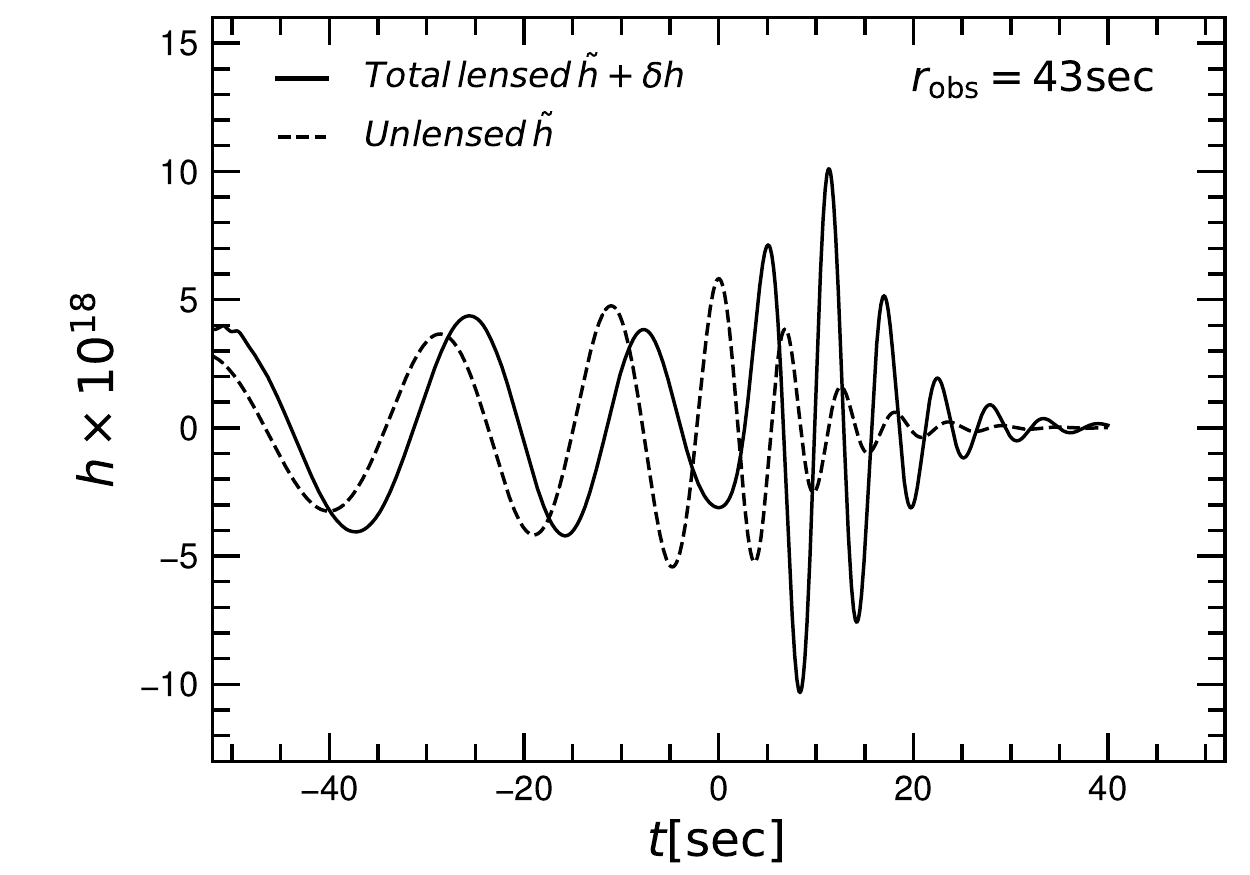}}
\caption{The total lensed waves $h =\delta h + \tilde{h}$ at the observer as a function of time (solid line). Compared with the unlensed original incident waves (dashed line), the waveforms of the total lensed waves change dramatically not only in the amplitude but also in the phase and pattern, especially near the merger of two black holes. \label{totalwave}}
\end{figure}

Figure~\ref{totalwave} shows the total lensed waves $h =\delta h + \tilde{h}$ at the observer as a function of time (solid line). Compared with the unlensed original incident waves (dashed line), the waveforms of the total lensed waves change dramatically not only in the amplitude but also in the phase and pattern. In the low-frequency regime, the amplitude of the total lensed waveform is similar to those of the incident waves. In the high-frequency regime, due to the lensing effect, the amplitude is much higher than those of the incident waves and the phase-shift is stronger as well due to the Shapiro time-delay.    

{\bf Conclusions.} 
We present the first numerical simulations of GWs passing through a potential well generated by a massive compact object. Unlike the previous analyses in the frequency-domain, which implicitly involve assumptions on the future behaviors of GWs at infinity, our simulations are in the time-domain. An advantage of the time-domain is that the propagation of GWs is local and is a well-posed "initial-value" problem for the hyperbolic equations with rigorous rooting in mathematics. Moreover, the wave equations are explicitly related to the well established physical laws, such as the conservation of energy-momentum. Our simulations, therefore, do not involve implicit unphysical assumptions and are solely based on first-principles calculations. Moreover, on the technical side, our simulations adopt the finite element method, which has a rigorous theory for numerical error control. Our simulations also adopt a symplectic scheme for the time evolution, which is inherently energy-preserving. Our numerical results, therefore, are robust (see our companion paper~\cite{He:2019orl} for details).

Based on these simulations, we investigate, for the first time, the impact of the locality and wavefront geometry on the waveforms of GWs in 3-D space when they travel in a potential well. We find that, contrary to the claim in Ref.~\cite{Suyama:2020lbf}, diffraction and the geometry of wavefront have dramatic effects on the speed of waves. The stronger the diffraction, the more easily the scattered waves along a faster ray can spread over the slower ones, which boosts the speed of the slower rays. The amounts of boost strongly depend on the wavelength. We find that in low-frequency regimes, the boost of wave speed is significant, and the shift on the phase is smaller. However, in the high-frequency regimes, the boost is less significant and the phase-shift is close to the maximum delay predicted by the Shapiro time-delay in the geometric limit. In addition to the wave speed, we also investigated the interference between the incident and the scattered waves. Unlike the previous analyses (e.g. Ref.~\cite{Takahashi_2017}) that only consider the scattered waves (see the explanations in Ref.~\cite{He:2019orl}), we find that such interference makes the total lensed waveforms dramatically different from those of the original incident waveforms (see Fig.~\ref{totalwave}) not only in the amplitude but also in the phase and pattern, especially near the merger of the two back-holes. This unique feature makes it much easier to be identified in future observations, such as eLISA~\cite{eLISA}, DECIGO~\cite{Sato_2009}, and Pulsar Timing Arrays (PTA)~\cite{2010CQGra..27h4013H}.

\section*{Acknowledgement} 
J.H.H. acknowledges support of Nanjing University and part of this work used the DiRAC@Durham facility managed by the Institute for Computational Cosmology on behalf of the STFC DiRAC HPC Facility (www.dirac.ac.uk). The equipment was funded by BEIS capital funding via STFC capital grants ST/K00042X/1, ST/P002293/1, ST/R002371/1 and ST/S002502/1, Durham University and STFC operations grant ST/R000832/1. DiRAC is part of the National e-Infrastructure.
\bibliography{myref}

\end{document}